\documentclass[12pt,a4paper]{iopart} 
\usepackage{iopams}
\usepackage{setstack,cite}
\usepackage{amssymb,graphicx}

\newcommand{\bea}{\begin{eqnarray}}
\newcommand{\eea}{\end{eqnarray}}
\newcommand{\ds}{\displaystyle}

\begin{document}
\bibliographystyle{revtex}
\title{Coherently coupled bright optical solitons and their collisions}
\author{T. Kanna$^1$, M. Vijayajayanthi$^2$, and M. Lakshmanan$^2$}
\address{$^1$ Post-Graduate and Research Department of Physics,  Bishop Heber College,\\Tiruchirapalli--620 017, India\\}
\address{
$^2$ Centre for Nonlinear Dynamics, School of Physics, Bharathidasan University, Tiruchirapalli--620 024, India}
\ead{kanna\_phy@bhc.edu.in(corresponding author)}
\ead{lakshman@cnld.bdu.ac.in}

\begin{abstract}
We obtain explicit bright one- and two-soliton solutions of the integrable case of the coherently coupled nonlinear Schr{\"o}dinger equations by applying a non-standard form of the Hirota's direct method. We find that the system admits both degenerate and non-degenerate solitons in which the latter can take single hump, double hump, and flat-top profiles. Our study on the collision dynamics of solitons in the integrable case shows that the collision among degenerate solitons and also the collision of non-degenerate solitons  are always standard elastic collisions. But the collision of a degenerate soliton with a non-degenerate soliton induces switching in the latter leaving the former unaffected after collision, thereby showing a different mechanism from that of the Manakov system.
\end{abstract}

\maketitle
\section{Introduction}
\indent Recently there has been considerable interest in studying the dynamics of multicomponent solitons/solitary waves in view of their wide range of applications encompassing science and engineering \cite{{r1},{r2}}. In the context of nonlinear optics, simultaneous propagation of multiple optical pulses or beams in nonlinear media is governed by a class of multicomponent nonlinear Schr\"odinger (NLS) type equations which is non-integrable in general.  These multicomponent NLS equations fall into two categories, namely incoherently coupled NLS equations and coherently coupled NLS equations \cite{r1}.  The integrable as well as non-integrable incoherently coupled NLS equations have been well studied in the literature \cite{{r1},{r2a}}.  Particularly, the studies on integrable Manakov system \cite{r3}, a two component nonlinear system with incoherent coupling, and also its integrable N-component generalization \cite{{r3},{r4},{r5},{r6}}, have revealed the fact that the bright solitons of these systems exhibit interesting collision scenario which is not possible in their single component counterparts.  This collision behaviour has been exploited in the construction of logic gates based on optical soliton collisions \cite{{r7},{r8}} and also such collisions lead to the possibility of multi-state logic \cite{{r6},{r6aa}}.

The set of coherently coupled NLS systems is another interesting class of nonlinear evolution equations for which much attention is yet to be paid.  The term coherent coupling here stands for the dependence of coupling on relative phases of the interacting fields.  A fairly general governing equation for coherently coupled orthogonally polarized waveguide modes in the Kerr medium (see for example, Sec. 9.4.1 in ref. \cite{r1}) is
\bea
i {q}_{1,Z}+\delta {q}_{1,TT}-\mu {q}_{1}+ (|{q}_{1}|^2+\sigma |{q}_{2}|^2){q}_{1}+\lambda {q}_2^2 {q}_{1}^*=0,\nonumber\\
i {q}_{2,Z}+\delta {q}_{2,TT}+\mu {q}_{2}+ (\sigma |{q}_{1}|^2+|{q}_{2}|^2){q}_2+\lambda {q}_{1}^2 {q}_{2}^*=0,
\label{eqn1}
\eea
where $Z$ and $T$ are the propagation direction and the transverse direction, respectively, ${q}_1$ and ${q}_2$ are slowly varying complex amplitudes in each polarization mode, $\mu$ is the degree of birefringence, $\sigma$ is the incoherent coupling parameter and $\lambda$ is the coherent coupling parameter.  Similar equations also arise in the context of short pulse propagation in weakly birefringent Kerr type nonlinear media \cite{{r1},{r2}}, where the co-ordinate $T$ corresponds to retarded time. In general the system (\ref{eqn1}) is non-integrable. An integrable non-dimensional coherently coupled NLS equation closely associated with equation (\ref{eqn1}) can be written as
\bea
i q_{1z}- q_{1tt}-\gamma (|q_1|^2+2 |q_2|^2)q_1-{\gamma} q_2^2 q_1^*=0,\nonumber\\
i q_{2z}- q_{2tt}-\gamma (2|q_1|^2+ |q_2|^2)q_2-\gamma q_1^2 q_2^*=0.
\label{eqn2}
\eea
The above set of equations results from equation (\ref{eqn1}) for the choice $\mu=0$ (low birefringence limit), $\sigma=2$, $\lambda=1$, a choice which is possible in a cubic anisotropic nonlinear medium where the parameters $\lambda$ and $\sigma$ can be chosen separately but their ratio is fixed as $\frac{\sigma}{\lambda}=2$  \cite{{r1},{r2}} and by performing the transformations $T\rightarrow \sqrt{\gamma \delta}~t$ and $Z\rightarrow -\gamma z$, where $\gamma>0$. Although the physical conditions for the above choice are stringent to obtain, we hope the exact results reported in this paper will serve as potential candidates in further analysis of the non-integrable coherently coupled NLS equations (\ref{eqn1}), which have received attention recently \cite{{r9},{r10},{r11},{r12}}.
\\
\indent Motivated by the above considerations, in this paper we have obtained  general soliton solutions of system (\ref{eqn2}) by applying a non-conventional form of Hirota's bilinearization method  \cite{r13}. Another integrable equation which can also be obtained from equation (\ref{eqn1}), having a form similar to equation (\ref{eqn2}) but with the replacement of the `$-$' sign appearing before the coherent coupling term by a `$+$' sign with $\gamma=1$, has been studied in refs. \cite{{r9},{r10}} and  special one- and two-soliton solutions with less number of parameters have been obtained by applying the Hirota's direct method. In fact, while obtaining the two-soliton solution by a linear superposition as reported in ref. \cite{r10}, following the lines of ref. \cite{r9}, the governing equation gets decoupled into two independent NLS equations and so the information regarding the coherent and incoherent coupling terms gets lost. This system can also be studied by applying a similar method as developed here and the results will be published separately. The main objective of this paper is to obtain an appropriate bilinear form of equation (\ref{eqn2}) resulting in more general soliton solutions as done in ref. \cite{r14} for the Sasa-Satsuma higher order NLS system. We also wish to investigate the soliton formation and propagation due to the combined effects of self phase modulation (SPM), cross phase modulation (XPM) and coherent coupling between the copropagating fields. Our study shows that there exist two distinct type of solitons, namely degenerate and non-degenerate solitons, where the non-degenerate solitons can have single and double hump profiles. Their collision behaviour is also fascinating. Particularly, the collision between degenerate and non-degenerate solitons shows a different kind of switching mechanism in the two component system (\ref{eqn2}) from that of the shape changing collisions occurring in the Manakov system \cite{{r4},{r6}}.

This paper is organized in the following manner. The non-standard way of obtaining the bilinear equations of the integrable system (\ref{eqn2}) by introducing an auxiliary function is discussed in section 2. The general one-soliton solution is obtained in section 3 and the degenerate and non-degenerate solitons are discussed. In section 4, the more general two-soliton solution reflecting the effects of coherent coupling terms during collision is obtained. The collisions of degenerate solitons and non-degenerate solitons are discussed separately in section 5 and  we have also analysed the collision of a degenerate soliton with a non-degenerate soliton in the same section. Final section 6 is allotted for conclusion.

\section{A non-standard bilinearization method for the coherently coupled NLS system}
\indent The soliton solutions of system (\ref{eqn2}) can be obtained by applying the  Hirota's bilinearization method \cite{r13}, which is a powerful tool for integrable nonlinear partial differential equations. To obtain the correct bilinear equations, resulting in more general soliton solutions displaying the effects of SPM, XPM, and coherent coupling, we adopt a non-standard method by introducing an auxiliary function, similar to the technique followed by Gilson {\it et al} \cite{r14} for the higher order NLS system.  By performing the bilinearizing transformation
\bea
q_1=\frac{g}{f}\;\; \mbox{and}\;\; q_2=\frac{h}{f},
\eea
to equation (\ref{eqn2}) and introducing an auxiliary function $s$, we obtain the following set of bilinear equations,
\numparts\bea
&&D_1~g\cdot f=-\gamma s g^*,\quad  D_1~h\cdot f=\gamma s h^*,\\
&&D_2~f \cdot f=2 \gamma \left(|g|^2+|h|^2\right),\quad sf=g^2- h^2,
\eea\label{beq}\endnumparts
where $D_1=iD_z-D_t^2$, $D_2=D_t^2$, $g$ and $h$ are complex functions, while $f$ is a real function,  $*$ denotes the complex conjugate and the Hirota's bilinear operators $D_z$ and $D_t$ are defined as \cite{r13}
\bea
D_z^{p}D_t^{q}(a\cdot b) =\bigg(\frac{\partial}{\partial z}-\frac{\partial}{\partial z'}\bigg)^p\bigg(\frac{\partial}{\partial t}-\frac{\partial}{\partial t'}\bigg)^q a(z,t)b(z',t')|_{ \ds (z=z', t=t')}.
\eea
\indent Note that the necessity for the introduction of an auxiliary function $s(z,t)$ becomes crucial as otherwise in the absence of $s$ in equation (4), only special cases of even one-soliton solution reported below will be obtained and for higher order solitons severe constraints on the soliton parameters will arise. The above set of equations (4) can be solved by introducing the following power series expansions for $g$, $h$, $f$, and $s$:
\numparts\bea
&&g=\chi g_1+\chi^3 g_3+\ldots, \quad h=\chi h_1+\chi^3 h_3+\ldots, \\
&&f=1+\chi^2 f_2+\chi^4 f_4+\ldots,\quad s=\chi^2 s_2+\chi^4 s_4+\ldots,
\eea\label{poweq}\endnumparts
where $\chi$ is the formal power series expansion parameter.  The resulting set of linear partial differential equations, after collecting the terms with the same powers in $\chi$, can be solved recursively to obtain the forms of $g$, $h$, $f$, and $s$.

\section{Bright one-soliton solutions}
\indent In order to obtain the one-soliton solution, unlike in the Manakov case \cite{{r4},{r5}}, here we restrict the power series expansion (6) as $g=\chi g_1+\chi^3 g_3$, $h=\chi h_1+\chi^3 h_3$, $f=1+\chi^2 f_2+\chi^4 f_4$, $s=\chi^2 s_2$.
After introducing this series expansion in equation (4) and by solving the resulting set of linear partial differential equations recursively, one can obtain the explicit one-soliton solution as
\numparts\bea
q_1=\frac{\alpha_1 e^{\eta_1}+e^{2\eta_1+\eta_1^*+\delta_{11}}}{1+e^{\eta_1+\eta_1^*+R_1}+e^{2\eta_1+2\eta_1^*+\epsilon_{11}}},\\
q_2=\frac{\beta_1 e^{\eta_1}+e^{2\eta_1+\eta_1^*+\rho_{11}}}{1+e^{\eta_1+\eta_1^*+R_1}+e^{2\eta_1+2\eta_1^*+\epsilon_{11}}},
\eea
where the auxiliary function takes the form 
\bea
&&s=(\alpha_1^2-\beta_1^2) e^{2\eta_1}.\\
\hspace{-2.5cm} \mbox{Here} \nonumber\\
&&\eta_1=k_1(t-i k_1 z),\quad
e^{\delta_{11}}=\frac{\gamma \alpha_1^* (\alpha_1^2-\beta_1^2)}{2 (k_1+k_1^*)^2},\quad e^{\rho_{11}}=\frac{-\gamma \beta_1^* (\alpha_1^2-\beta_1^2)}{2 (k_1+k_1^*)^2}, \\
&&e^{R_1}=\frac{\gamma (|\alpha_1|^2+|\beta_1|^2)}{(k_1+k_1^*)^2},\quad
e^{\epsilon_{11}}=\frac{\gamma^2(\alpha_1^2-\beta_1^2) (\alpha_1^{*2}-\beta_1^{*2})}{4 (k_1+k_1^*)^4}.
\eea\label{onesol}\endnumparts
\noindent\underline{Case(i): $\alpha_1^2-\beta_1^2=0$}\\
\indent This choice $\alpha_1^2-\beta_1^2=0$ always results in the standard ``sech" profile for the bright soliton solution (7).  It can be expressed as
\bea
q_1&=&\left(\frac{\alpha_1}{2}e^{-\frac{R_1}{2}}\right)\mbox{sech}\left(\eta_{1R}+\frac{R_1}{2}
\right)e^{i\eta_{1I}} \equiv A_1 \mbox{sech}\left(\eta_{1R}+\frac{R_1}{2}\right)e^{i\eta_{1I}},
\label{ms}
\eea
and $q_2=\pm q_1$ corresponding to $\beta_1=\pm \alpha_1$ so that $|q_1|^2=|q_2|^2$. Here $A_1=\left(\frac{\alpha_1}{2}e^{-\frac{R_1}{2}}\right)$, $R_1=\mbox{log}\left(\frac{2\gamma|\alpha_1|^2}{(k_1+k_1^*)^2}\right)$,  and $\eta_1=\eta_{1R}+i\eta_{1I}$, where $\eta_{1R}=k_{1R}(t+2k_{1I}z)$ and $\eta_{1I}=k_{1I}{t}+(k_{1I}^2-k_{1R}^2){z}$. Throughout this paper the subscripts $R$ and $I$ represent the real and imaginary parts, respectively. We call the solitons arising for the choice $\alpha_1^2-\beta_1^2=0$ as \emph{degenerate} solitons, owing to the fact that such solitons posses the same intensity profile in both the components $q_1$ and $q_2$ and are characterized by two complex parameters $\alpha_1$ and $k_1$ or four real parameters, instead of the six real parameters in the Manakov case \cite{r6}. Here $A_1$, $-2k_{1I}$, and $\frac{R_1}{2k_{1R}}$ are the amplitude, velocity, and central position of the soliton, respectively. Note that $\frac{A_1}{k_{1R}}$ is related to the polarization of the pulse/beam.  The degenerate soliton having a single hump profile is depicted in figure 1 for the parameters $\gamma=2$, $k_1=1+i$, and $\alpha_1=\beta_1=1$ at $t=0 $.
\begin{figure}[h]
\centering
\includegraphics[width=0.63\linewidth]{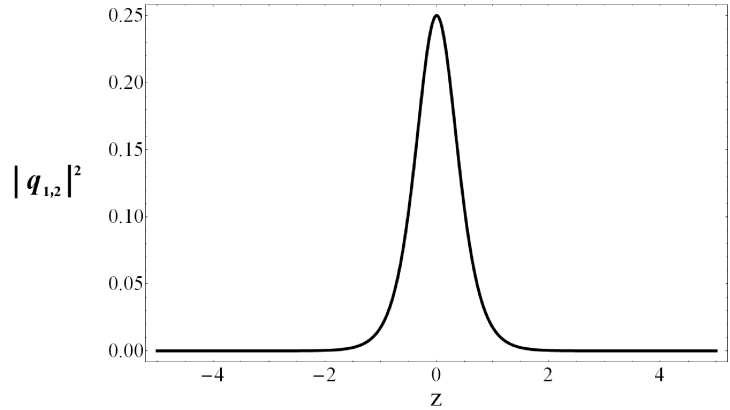}
\caption{Degenerate one-soliton at $t=0$ (parameters are as given in the text). }
\end{figure}\\
\indent Note that in the present case, since $\alpha_1^2-\beta_1^2=0$, the auxiliary function $s$ vanishes, see equation (7c), and so from the bilinear equations (4) one can easily infer that the effect of coherent coupling vanishes.\\
\noindent\underline{Case(ii): $\alpha_1^2-\beta_1^2\neq0$}\\
\indent The nature of soliton for the other choice $\alpha_1^2-\beta_1^2\neq0$, can be understood by rewriting $q_1$ and $q_2$ in the expression (7) as
\bea
\hspace{-1.5cm} q_j=\frac{2A_j\left[\mbox{cos}(P_j)\mbox{cosh}\left(\eta_{1R}+\frac{\epsilon_{11}}{4}\right)
+i\;\mbox{sin}(P_j)\mbox{sinh}\left(\eta_{1R}+\frac{\epsilon_{11}}{4}\right) \right]e^{i\eta_{1I}}}{4 \mbox{cosh}^2\left(\eta_{1R}+\frac{\epsilon_{11}}{4}\right)+L}, \quad j=1,2,~~~~
\label{os}
\eea
where $A_1=e^{\left({\frac{l_1+\delta_{11}-\epsilon_{11}}{2}}\right)}$, $A_2=e^{\left({\frac{l_2+\rho_{11}-\epsilon_{11}}{2}}\right)}$, $P_1=\frac{(\delta_{11I}-l_{1I})}{2}$, $P_2=\frac{(\rho_{11I}-l_{2I})}{2}$, $L=e^{\left({R_1-\frac{\epsilon_{11}}{2}}\right)}-2$, $\eta_{1R}=k_{1R}(t+2k_{1I}z)$, $\eta_{1I}=k_{1I}{t}+(k_{1I}^2-k_{1R}^2){z}$, $l_{1}=\ln({\alpha_1})$, and $l_{2}=\ln({\beta_1})$. Also, the quantities $\delta_{11}$, $\rho_{11}$, $R_1$, and $\epsilon_{11}$ are as defined in equation (7).
Here $A_j$ represents the amplitude of the soliton in the $j$-th component and for this case by the term amplitude we mean the peak value of the soliton profile. The speed of the soliton is given by $2k_{1I}$ and its central position is $\frac{\epsilon_{11}}{4k_{1R}}$.  It can be noticed that by rewriting the expression (9) for the choice $\alpha_1=\beta_1$, it reduces to equation (8). The general form presented here will be of use in the asymptotic analysis of the two-soliton and multi-soliton solutions.  We refer to the above soliton as \emph{non-degenerate} due to their distinct intensity profiles in the $q_1$ and $q_2$ components. In contrast to the degenerate solitons these solitons can vary their profile from a single hump to a double hump through a flat-top profile as the parameters are varied. A double hump soliton and a flat-top soliton appearing in the $q_1$ and $q_2$ components, respectively, at $t=0 $ are shown in figure 2 for the parameters $\gamma=2$, $k_1=1+i$, $\alpha_1=0.7114$, and $\beta_1=1$. 
\begin{figure}[h]
\centering
\includegraphics[width=0.6\linewidth]{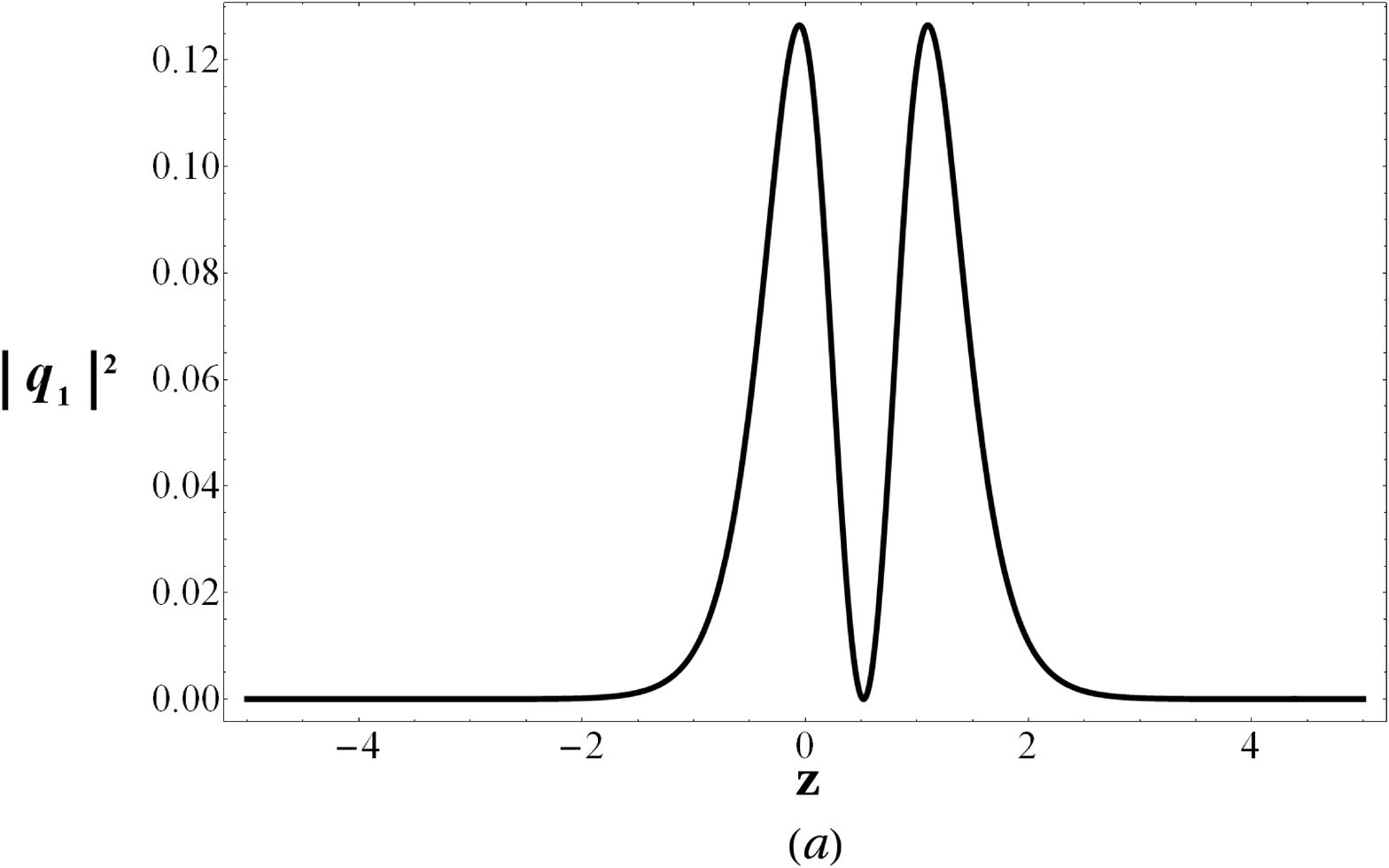}~\\~\includegraphics[width=0.6\linewidth]{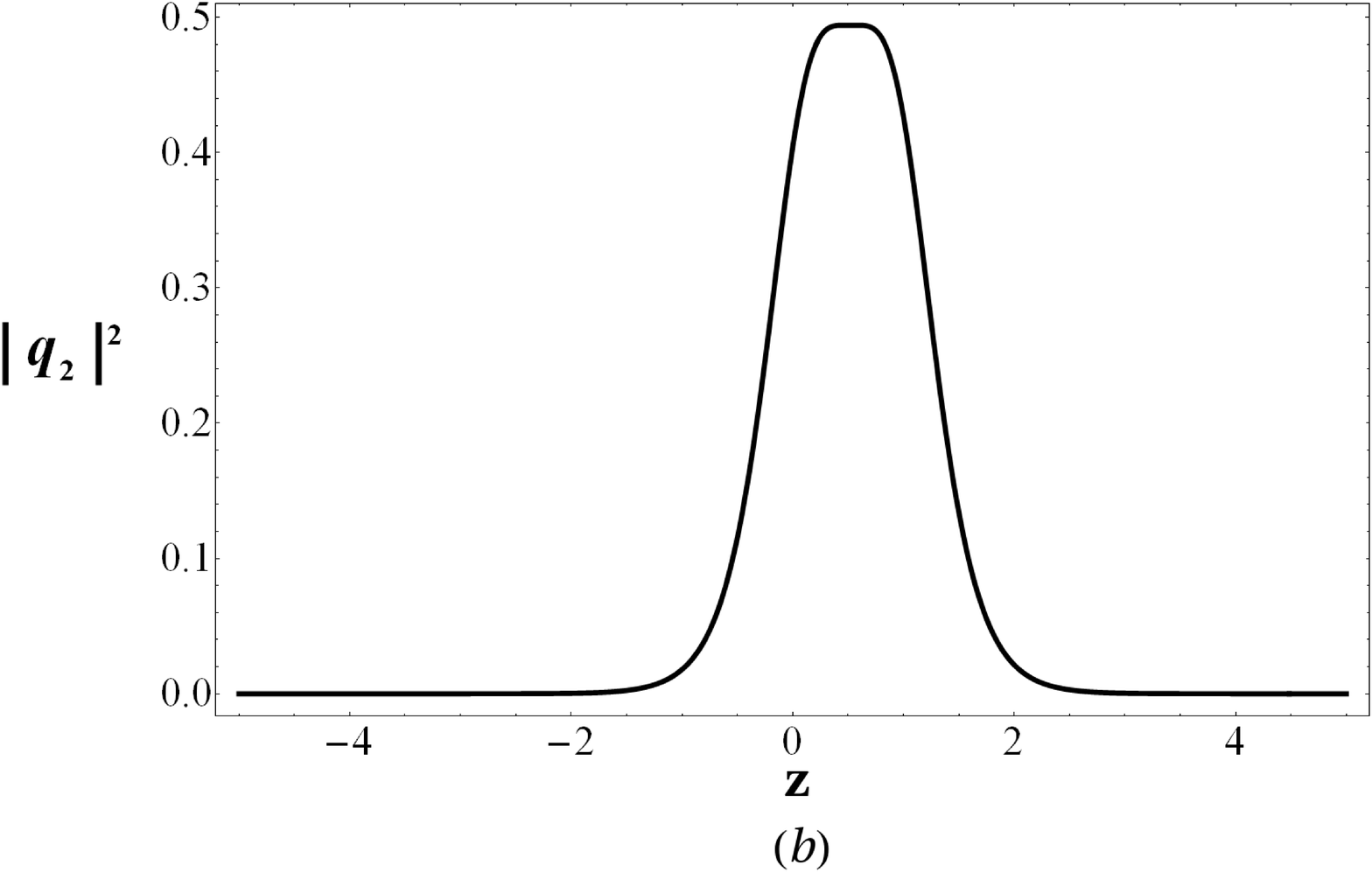}
\caption{A non-degenerate soliton at $t=0$: (a) Double hump non-degenerate soliton in the $q_1$ component. (b) Flat-top non-degenerate soliton in the $q_2$ component (parameters are as given in the text).}
\end{figure}
Similar kind of flat-top structures have been reported in complex Ginzburg-Landau equation \cite{r2}. Note that in equation (9) also the standard $\mbox{sech}$ type soliton occurs for a particular choice of the parameters, namely $\alpha_1 \beta_1^*+\alpha_1^* \beta_1=0$, for which $P_1$,  $P_2$, and $L$ become  zero in equation (9). However, in the present case, the effect of coherent coupling does not vanish unlike the case of degenerate solitons.\\
\indent It can be noticed that equation (\ref{eqn2}) is embedded into the matrix NLS equation which is integrable via Inverse Scattering Transform (IST) method \cite{{r15},{r16}}. Cases (i) and (ii) have also been reported in refs. \cite{{r15},{r16}} for a three component version of the equation considered in this paper by applying the results of the IST method for the matrix NLS equation and the corresponding solutions were referred as ferromagnetic and polar solitons, respectively, in the context of multicomponent spinor condensates.  Here we have obtained similar kind of more general soliton solutions for the two component case itself by applying a non-standard type of Hirota's bilinearization method.

\section{Bright two-soliton solution}
\indent The two-soliton solution of the system (\ref{eqn2}) can be obtained after terminating the power series (6) as $g=\chi g_1+\chi^3 g_3+\chi^5 g_5+\chi^7 g_7$, $h=\chi h_1+\chi^3 h_3+\chi^5 h_5+\chi^7 h_7$,
$f=1+\chi^2 f_2+\chi^4 f_4+\chi^6 f_6+\chi^8 f_8$, $s=\chi^2 s_2+\chi^4 s_4+\chi^6 s_6$ and again by solving the resultant linear partial differential equations recursively.  Then the explicit form of the two-soliton solution can be written as
\numparts\bea
q_j&=&\frac{N^{(j)}}{D},\;\; \quad\;\;j=1,2.
\eea
The functions $N^{(1)}$, $N^{(2)}$ and $D$ in (10a) are given by the expressions
\bea
N^{(1)}=&&\alpha _1 e^{\eta _1} +\alpha _2 e^{\eta _2}+e^{2 \eta _1+\eta _1^{*}+\delta_{11}}+e^{2 \eta _1+\eta _2^{*}+\delta_{12}}+e^{2 \eta _2+\eta _1^{*}+\delta_{21}}+e^{2 \eta _2+\eta _2^{*}+\delta_{22}}\nonumber\\
&&+e^{\eta_1+\eta_1^*+\eta_2+\delta_1}+e^{\eta_2+\eta_2^*+\eta_1+\delta_2}+e^{2\eta_1+2\eta_1^*+\eta_2+\mu_{11}}
+e^{2\eta_1+2\eta_2^*+\eta_2+\mu_{12}}\nonumber\\&&+e^{2\eta_2+2\eta_1^*+\eta_1+\mu_{21}}+e^{2\eta_2+2\eta_2^*+\eta_1+\mu_{22}}
+e^{2\eta_1+\eta_1^*+\eta_2+\eta_2^*+\mu_1}\nonumber\\&&+e^{2\eta_2+\eta_2^*+\eta_1+\eta_1^*+\mu_2}
+e^{2\eta_1+2\eta_1^*+2\eta_2+\eta_2^*+\phi_1}+e^{2\eta_1+2\eta_2+2\eta_2^*+\eta_1^*+\phi_2},~~~~~~~\\
N^{(2)}=&&\beta _1 e^{\eta_1} +\beta_2 e^{\eta_2}+e^{2 \eta_1+\eta _1^{*}+\rho_{11}}+e^{2 \eta_1+\eta_2^{*}+\rho_{12}}+e^{2 \eta _2+\eta _1^{*}+\rho_{21}}+e^{2 \eta _2+\eta _2^{*}+\rho_{22}}\nonumber\\&&+e^{\eta_1+\eta_1^*+\eta_2+\rho_1}+e^{\eta_2+\eta_2^*+\eta_1+\rho_2}
+e^{2\eta_1+2\eta_1^*+\eta_2+\nu_{11}}+e^{2\eta_1+2\eta_2^*+\eta_2+\nu_{12}}\nonumber\\
&&+e^{2\eta_2+2\eta_1^*+\eta_1+\nu_{21}}+e^{2\eta_2+2\eta_2^*+\eta_1+\nu_{22}}+e^{2\eta_1+\eta_1^*+\eta_2+\eta_2^*+\nu_1}\nonumber\\
&&+e^{2\eta_2+\eta_2^*+\eta_1+\eta_1^*+\nu_2}+e^{2\eta_1+2\eta_1^*+2\eta_2+\eta_2^*+\psi_1}+e^{2\eta_1+2\eta_2+2\eta_2^*+\eta_1^*+\psi_2},~~~~\\
~~~D=&&1+e^{\eta_1+\eta_1^*+R_1}+e^{\eta_1+\eta_2^*+\delta_0}+e^{\eta_2+\eta_1^*+\delta_0^*}+e^{\eta_2+\eta_2^*+R_2}
+e^{2\eta_1+2\eta_1^*+\epsilon_{11}}\nonumber\\&&+e^{2\eta_1+2\eta_2^*+\epsilon_{12}}+e^{2\eta_2+2\eta_1^*+\epsilon_{21}}
+e^{2\eta_2+2\eta_2^*+\epsilon_{22}}+e^{2\eta_1+\eta_1^*+\eta_2^*+\tau_1}\nonumber\\
&&+e^{2\eta_1^*+\eta_1+\eta_2+\tau_1^*}+e^{2\eta_2+\eta_1^*+\eta_2^*+\tau_2}+e^{2\eta_2^*+\eta_1+\eta_2+\tau_2^*}
+e^{\eta_1+\eta_1^*+\eta_2+\eta_2^*+R_3}\nonumber\\
&&+e^{2\eta_1+2\eta_1^*+\eta_2+\eta_2^*+\theta_{11}}+e^{2\eta_1+2\eta_2^*+\eta_2+\eta_1^*+\theta_{12}}
+e^{2\eta_2+2\eta_1^*+\eta_1+\eta_2^*+\theta_{21}}\nonumber\\&&+e^{2\eta_2+2\eta_2^*+\eta_1+\eta_1^*+\theta_{22}}
+e^{2(\eta_1+\eta_1^*+\eta_2+\eta_2^*)+R_4},~~~~
\eea
and the auxiliary  function $s$ is determined as
\bea
s=&&(\alpha _1^2-\beta _1^2)e^{2 \eta _1}+(\alpha _2^2-\beta _2^2)e^{2 \eta _2}+2(\alpha _1\alpha _2-\beta_1\beta _2)e^{\eta _1+\eta _2}+e^{\eta _1+\eta _1^*+2\eta _2+\lambda _{11}}\nonumber\\
&&+e^{\eta _1+\eta _2^*+2 \eta _2+\lambda _{12}}+e^{\eta _2+\eta _1^*+2 \eta_1+\lambda _{21}}+e^{\eta _2+\eta _2^*+2 \eta _1+\lambda _{22}}+e^{2 \eta _1+2 \eta _1^*+2\eta _2+\lambda _1} \nonumber\\&&+e^{2 \eta _1+2\eta _2+2 \eta _2^*+\lambda _2}+e^{2\eta _1+\eta_1^*+2 \eta _2+\eta _2^*+\lambda _3}. \label{eqn2s}
\eea
\endnumparts
Here $\eta_i=k_i(t-i k_iz)$, $i=1,2$. The real and imaginary parts of $\eta_j$ are given by $\eta_{jR}=k_{jR}({t}+2k_{jI}{z})$ and $\eta_{jI}=k_{jI}{t}+(k_{jI}^2-k_{jR}^2){z}$, $j=1,2$. Various quantities appearing in equation (10) are given in the Appendix, as they are rather lengthy expressions. In order to understand the structure of the above two-soliton solution, we now perform an asymptotic analysis and analyse the nature of the soliton collisions in the present system.

\section{Collision of solitons}
\indent The two-soliton solution obtained in the previous section represents the interaction of two solitons. It is of interest to consider the collision among non-degenerate solitons and degenerate solitons, and also the collision between the non-degenerate and degenerate solitons. For this purpose we perform the asymptotic analysis of the two-soliton solution (10) by considering the case where $k_{1R}, k_{2R} > 0$ and $k_{1I} > k_{2I}$, without loss of generality. The analysis is straightforward for the other choices of $k_{jR}$ and $k_{jI}$, $j=1,2$.
\subsection{Collision of non-degenerate solitons ($\alpha_j^2\neq\beta_j^2,~j=1,2$)}
\indent The asymptotic forms of $S_1$ and $S_2$ before collision ($z\rightarrow-\infty$) and after collision ($z\rightarrow+\infty$) can be deduced from equation (10) as follows. The quantities $\eta_{jR}$ and $\eta_{jI}$, $j=1,2$, appearing in the following asymptotic expressions are defined below equation (10).\\
{\it 1. Before Collision ($z\rightarrow-\infty$) }\\
\noindent\underline{Soliton $S_1$} ($\eta_{1R}\simeq0, \eta_{2R}\rightarrow -\infty$):
\numparts
\bea
\hspace{-2.0cm} \left(\begin{array}{c}
  q_1^{1-} \\
  q_2^{1-}
\end{array}\right) \simeq \frac{1}{D_1}
    \left(\begin{array}{cc}
         A_1^{1-}&0 \\
        0&A_2^{1-}
     \end{array} \right)
     \left(\begin{array}{cc}
         \mbox{cos}(P_1)~~ & i~\mbox{sin}(P_1)\\
         \mbox{cos}(P_2)~~ & i~\mbox{sin}(P_2) \\
       \end{array}\right)
     \left(\begin{array}{c}
         \mbox{cosh}(\eta_{1R}^{-}) \\
         \mbox{sinh}(\eta_{1R}^{-}) \\
       \end{array}\right)e^{i\eta_{1I}},
\eea
\bea 
\hspace{-2.5cm} \mbox{where}\nonumber\\ 
&&\left(\begin{array}{c}
        			 A_1^{1-} \\
      				  A_2^{1-}
     			\end{array} \right)= 2e^{-\frac{(R_4+\epsilon_{22})}{2}}
				\left(\begin{array}{c}
       		  		e^{\frac{(\mu_{22}+\phi_{2})}{2}} \\
        			 e^{\frac{(\nu_{22}+\psi_{2})}{2}}
     			\end{array} \right),\\
&& D_1= 4~\mbox{cosh}^2(\eta_{1R}^{-})+e^{\left({\theta_{22}-\frac{(R_4+\epsilon_{22})}{2}}\right)}-2.
\eea 
\endnumparts In the above, $P_1=\frac{\phi_{2I}-\mu_{22I}}{2}, ~P_2=\frac{\psi_{2I}-\nu_{22I}}{2}$,
 and
$\eta_{1R}^{-}=\eta_{1R}+\frac{R_4-\epsilon_{22}}{4}$. 
Here and in the following the superscript denotes the soliton and the subscript denotes the component and - (+) sign appearing in the superscript represents the asymptotic form of the soliton before (after) interaction.\\
\noindent\underline{Soliton $S_2$} ($\eta_{2R}\simeq0, \eta_{1R}\rightarrow \infty$):
\numparts
\bea
\hspace{-2.0cm} \left(\begin{array}{c}
  q_1^{2-} \\
  q_2^{2-}
\end{array}\right) \simeq \frac{{1}}{D_2}
    \left(\begin{array}{cc}
         A_1^{2-} &0\\
        0&A_2^{2-}
     \end{array} \right)
     \left(\begin{array}{cc}
         \mbox{cos}(Q_1)~~ & i~\mbox{sin}(Q_1)\\
         \mbox{cos}(Q_2)~~ & i~\mbox{sin}(Q_2) \\
       \end{array}\right)
     \left(\begin{array}{c}
         \mbox{cosh}(\eta_{2R}^{-}) \\
         \mbox{sinh}(\eta_{2R}^{-}) \\
     \end{array}\right)e^{i\eta_{2I}},
\eea
\bea 
\hspace{-2.5cm} \mbox{where}\nonumber\\ 
&&\left(\begin{array}{c}
         A_1^{2-} \\
        A_2^{2-}
     \end{array} \right)= 2e^{-\frac{\epsilon_{22}}{2}}
     \left( \begin{array}{c}
       {e^{\frac{{l_1^-}+\delta_{22}}{2}}}\\
       {e^{\frac{{l_2^-}+\rho_{22}}{2}}} \\
       \end{array}\right), \\
       && D_2= 4~\mbox{cosh}^2(\eta_{2R}^{-})+e^{\left(R_2-\frac{\epsilon_{22}}{2}\right)}-2.\eea \endnumparts 
        Here, $Q_1=\frac{\delta_{22I}-l_{1I}^-}{2}$, $Q_2=\frac{\rho_{22I}-l_{2I}^-}{2}$,
$l_1^-=\mbox{ln}(\alpha_2)$, $l_2^-=\mbox{ln}(\beta_2)$,
 and
$\eta_{2R}^{-}=\eta_{2R}+\frac{\epsilon_{22}}{4}$. All the quantities appearing in the above asymptotic expressions (11) and (12) are defined in the Appendix.\\
\noindent {\it {2. After Collision:}}\\
\indent The asymptotic expressions after collision are similar to those of before collision expressions with the replacement of $A_l^{j-}$ and  $\eta_{jR}^-$ by $A_l^{j+}$ and  $\eta_{jR}^+$, respectively for the soliton $S_j$, $j=1,2$, where $A_l^{1+}=\frac{(k_1+k_2^*)(k_1^*-k_2^*)}{(k_1^*+k_2)(k_1-k_2)}A_l^{1-}$, $A_l^{2+}=\frac{(k_1+k_2^*)(k_1-k_2)}{(k_1^*+k_2)(k_1^*-k_2^*)}A_l^{2-}$, $l=1,2$, $\eta_{1R}^{+}=\eta_{1R}+\frac{\epsilon_{11}}{4}$, and $\eta_{2R}^{+}=\eta_{2R}+\frac{(R_4-\epsilon_{11})}{4}$. The quantities $\epsilon_{11}$ and $R_4$ are given in the Appendix. 
\begin{figure}[h]
\centering
\includegraphics[width=0.6\linewidth]{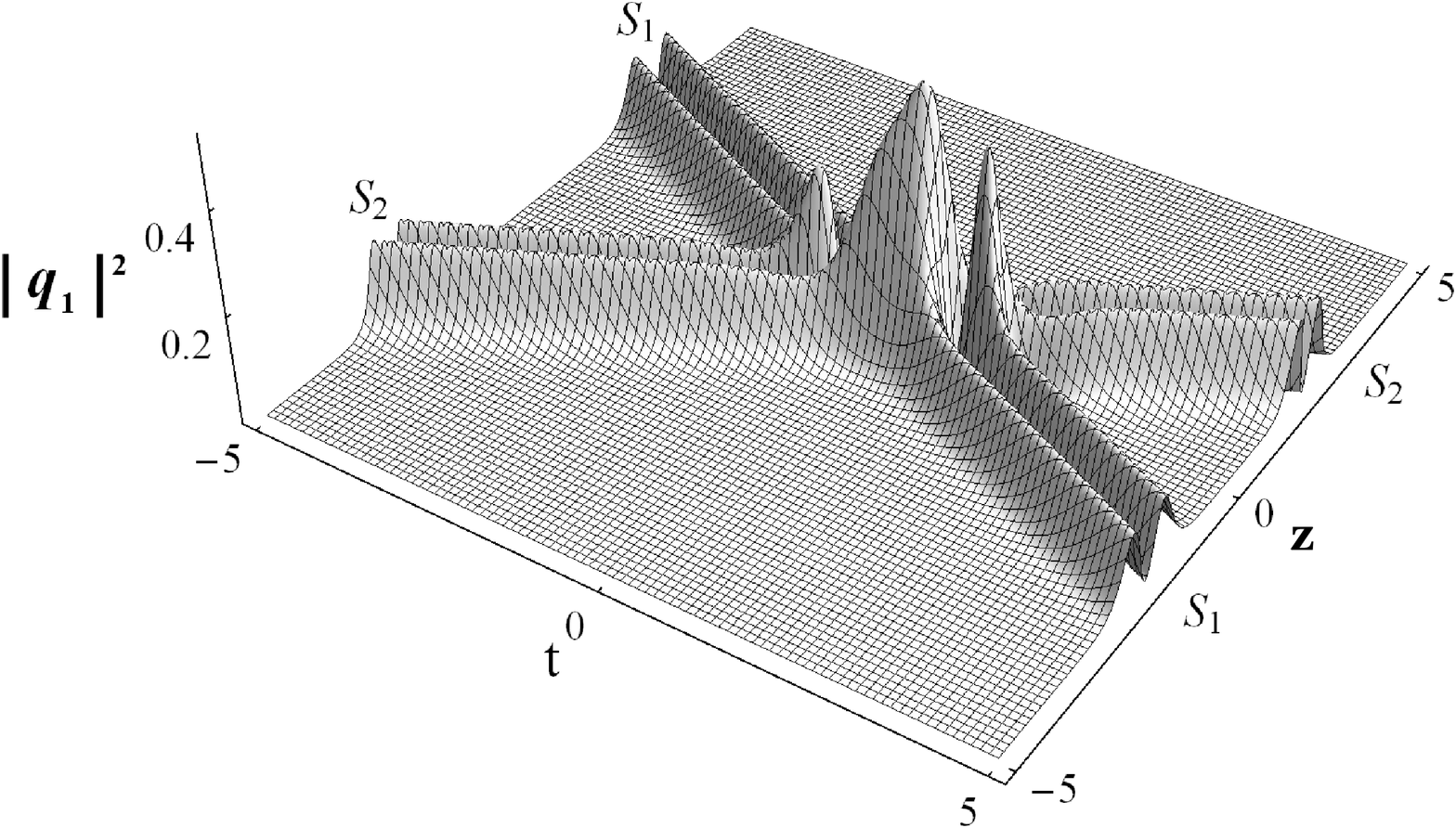}\\~~\includegraphics[width=0.6\linewidth]{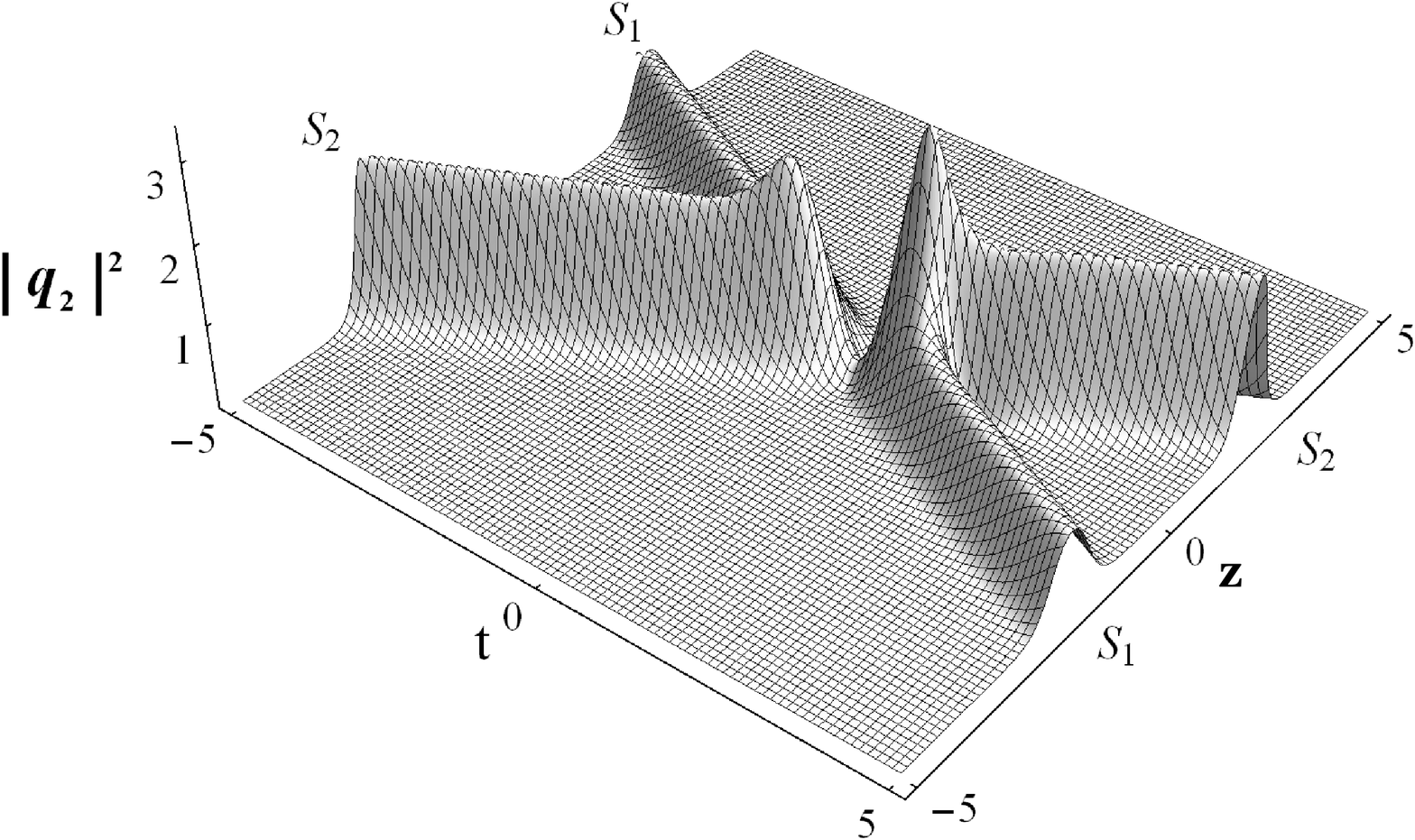}
\caption{Elastic collision of non-degenerate solitons (parameters are as given in the text).}
\end{figure}
One can easily check that  the intensities  before and after interaction are same, that is,
 $|A_l^{j-}|^2=|A_l^{j+}|^2$, $j, l=1,2$. Also, the velocities of the two colliding solitons $S_1$ and $S_2$ are exactly the same before and after collision except for a phase shift which is found to be $\Phi_1=\frac{\epsilon_{11}+\epsilon_{22}-R_4}{4k_{1R}}\equiv \frac{1}{k_{1R}} \mbox{ln}\left[\frac{(k_2+k_1^*)(k_1+k_2^*)}{(k_1-k_2)(k_1^*-k_2^*)}\right]$, for the soliton $S_1$ and the soliton $S_2$ experiences a phase shift $\Phi_2=-\Phi_1\left(\frac{k_{1R}}{k_{2R}}\right)$. Thus our analysis on the non-degenerate solitons arising for the general choice $\alpha_j^2-\beta_j^2\neq0$ shows that these type of solitons always undergo standard elastic collision in the coherently coupled NLS system (2) and one such collision is depicted in figure 3 for the parameters, $\gamma=3$, $k_1=1.5+i$, $k_2=2-i$, $\alpha_1=1$, $\beta_1=1.7$, $\alpha_2=1$, and $\beta_2=2$. In figure 3, the double hump solitons undergo elastic collision in the $q_1$ component and in the $q_2$ component the single hump solitons exhibit elastic collision. One can also have the double hump solitons in both the components, for suitable choices of parameters.

\subsection{Collision of degenerate solitons}
\indent The degenerate solitons arise for the choice $\alpha_j^2-\beta_j^2=0$, $j=1,2$. This happens when $\alpha_1=\pm\beta_1$ and $\alpha_2=\pm\beta_2$. In the following, we perform the analysis for the case $\alpha_1=\beta_1$ and $\alpha_2=\beta_2$. For the other choices, that is, $\alpha_1=\beta_1$ and $\alpha_2=-\beta_2$ or $\alpha_1=-\beta_1$ and $\alpha_2=\beta_2$, also the collision scenario is similar to the choice discussed in this subsection.\\
{\it {1. Before Collision ($z \rightarrow -\infty$)}}\\
\noindent\underline{Soliton $S_1$:}
\bea
q_1^{1-}&=&q_2^{1-}=A^{1-}\mbox{sech}(\eta_{1R}^-)e^{i\eta_{1I}},
\eea
where $
A^{1-}=\frac{e^{\delta_2-\left(\frac{R_2+R_3}{2}\right)}}{2}$ and $\eta_{1R}^-=\eta_{1R}+\frac{R_3-R_2}{2}$.

\noindent\underline{Soliton $S_2$:}
\bea
q_1^{2-}&=&q_2^{2-}=A^{2-}\mbox{sech}(\eta_{2R}^-)e^{i\eta_{2I}},
\eea
where $A^{2-}=\frac{\alpha_2}{2}e^{-\frac{R_2}{2}}$ and $\eta_{2R}^-=\eta_{2R}+\frac{R_2}{2}$.

\noindent {\it {2. After Collision ($z \rightarrow +\infty$)}}\\
\noindent\underline{Soliton $S_1$:}
\bea
q_1^{1+}=q_2^{1+}=A^{1+}\mbox{sech}(\eta_{1R}^+)e^{i\eta_{1I}},
\eea
where $A^{1+}=\frac{\alpha_1}{2}e^{-\frac{R_1}{2}}$ and $\eta_{1R}^+=\eta_{1R}+\frac{R_1}{2}$.\\
\noindent \underline{Soliton $S_2$:}
\bea
q_1^{2+}=q_2^{2+}=A^{2+}~\mbox{sech}(\eta_{2R}^+)e^{i\eta_{2I}},
\eea
where $A^{2+}=\frac{e^{\delta_1-\left(\frac{R_1+R_3}{2}\right)}}{2}$ and $\eta_{2R}^+=\eta_{2R}+\frac{R_3-R_1}{2}$.\\
\indent All the quantities appearing in the above expressions (13-16) can be obtained from the corresponding quantities defined in the Appendix with the substitution $\beta_j=\alpha_j$, $j=1,2$, and the real and imaginary parts of $\eta_{j}$-s are defined below equation (10). From the above expressions, one can show that the amplitudes $A^j$-s before and after the interaction are related through the  expressions $
A^{1+}=\left[\frac{(k_1^*-k_2^*)(k_1+k_2^*)}{(k_1-k_2)(k_1^*+k_2)}\right]A^{1-}\;\mbox{and}\;
A^{2+}=\left[\frac{(k_1-k_2)(k_1+k_2^*)}{(k_1^*-k_2^*)(k_1^*+k_2)}\right]A^{2-}$, which shows that the intensities before and after interactions are the same, that is $|A^{j+}|^2=|A^{j-}|^2, j=1,2$. Also the soliton $S_1$ undergoes a phase shift $\Phi_1=\frac{R_1+R_2-R_3}{2k_{1R}}$, whereas  the soliton $S_2$ experiences a phase shift $\Phi_2=-\Phi_1\left(\frac{k_{1R}}{k_{2R}}\right)$ during collision. Thus the degenerate solitons always undergo standard elastic collision as that of the NLS solitons.
\subsection{Collision between degenerate and non-degenerate solitons}
\indent  The collision of a degenerate soliton ($\alpha_j^2-\beta_j^2=0$) with a non-degenerate soliton ($\alpha_j^2-\beta_j^2\neq0$) exhibits very interesting collision properties. Here we consider the collision of a non-degenerate soliton $S_1$ ($\alpha_1\neq\beta_1$) with a degenerate soliton $S_2$ ($\alpha_2=\beta_2$). Note that the analysis can also be performed for the other possible choices like $\alpha_2=-\beta_2$, but here also one can infer the same kind of collision scenario as for the present choice $\alpha_2=\beta_2$. The asymptotic forms of the solitons $S_1$ and $S_2$ are presented below.\\
{\it {1. Before Collision}}\\
\noindent\underline{Soliton $S_1$:}
\bea
\fl\left(\begin{array}{c}
      q_1^{1-} \\
      q_2^{1-} \\
\end{array}\right)=\frac{1}{D^{1-}}
   \left(\begin{array}{cc}
      A_1^{1-} &0 \\
     0& A_2^{1-} \\
   \end{array}\right)
       \left(\begin{array}{cc}
           \mbox{cos}(P_1^-) & i\; \mbox{sin}(P_1^-) \\
           \mbox{cos}(Q_1^-) & i\; \mbox{sin}(Q_1^-) \\
       \end{array}\right) \left(\begin{array}{c}
         \mbox{cosh}(\eta_{1R}^{-}) \\
         \mbox{sinh}(\eta_{1R}^{-}) \\
       \end{array}\right)e^{i\eta_{1I}},
\eea
where 
$\left(\begin{array}{c}
      A_1^{1-} \\
      A_2^{1-} \\
 \end{array}\right)=2
   \left(\begin{array}{c}
     e^{\frac{\delta_2+\mu_1-\theta_{11}-R_2}{2}} \\
     e^{\frac{\rho_2+\nu_1-\theta_{11}-R_2}{2}} \\
   \end{array}\right)$, 
$D^{1-}=4\mbox{cosh}^2(\eta_{1R}^-)+L^{1-}$,
$P_1^-=\frac{\delta_{2I}-\mu_{1I}}{2}$, $Q_1^-=\frac{\rho_{2I}-\nu_{1I}}{2}$, $\eta_{1R}^-=\eta_{1R}+\frac{\theta_{11}-R_2}{4}$, and $L^{1-}=e^{\left({R_3}-\frac{(\theta_{11}+R_2)}{2}\right)}-2$. Note that the expressions for various quantities appearing in equation (17) and in the equations (18-20) given below can be obtained from the corresponding quantities defined in the Appendix by putting $\beta_2=\alpha_2$.

\noindent\underline{Soliton $S_2$:}
\bea
q_1^{2-}=q_2^{2-}=A^{2-} \mbox{sech}(\eta_{2R}^-) e^{i\eta_{2I}},
\eea
where $A^{2-}=\frac{\alpha_2}{2} e^{-\frac{R_2}{2}}$ and $\eta_{2R}^-=\eta_{2R}+\frac{R_2}{2}$.\\
{\it {2. After Collision}}\\
\noindent\underline{Soliton $S_1$:}
\bea
\fl\left(\begin{array}{c}
      q_1^{1+} \\
      q_2^{1+} \\
\end{array}\right)=\frac{1}{D^{1+}}
   \left(\begin{array}{cc}
      A_1^{1+}&0 \\
     0& A_2^{1+} \\
   \end{array}\right)
       \left(\begin{array}{cc}
           \mbox{cos}(P_1^+) & i\; \mbox{sin}(P_1^+) \\
           \mbox{cos}(Q_1^+) & i\; \mbox{sin}(Q_1^+) \\
       \end{array}\right) \left(\begin{array}{c}
         \mbox{cosh}(\eta_{1R}^{+}) \\
         \mbox{sinh}(\eta_{1R}^{+}) \\
       \end{array}\right)e^{i\eta_{1I}},
\eea
where $P_1^+=\frac{\delta_{11I}-{l_{1I}^+}}{2}$, $Q_1^+=\frac{\rho_{11I}-{l_{2I}^+}}{2}$,  $D^{1+}=4\mbox{cosh}^2(\eta_{1R}^+)+L^{1+}$, 
  ${l_{1I}^+}=\mbox{ln}(\alpha_1)$, ${l_{2I}^+}=\mbox{ln}(\beta_1)$, $\eta_{1R}^+=\eta_{1R}+\frac{\epsilon_{11}}{4}$, and $L^{1+}=e^{{R_1}-\frac{\epsilon_{11}}{2}}-2$. The amplitudes $A_1^{1+}$ and $A_2^{1+}$ are given by the relations $A_1^{1+}=T_1~A_1^{1-}$ and $A_2^{1+}=T_2~A_2^{1-}$. Here the transition amplitude 
\bea
T_1=\sqrt{\frac{4(k_1^*-k_2^*)^2(k_1+k_2^*)^2 \alpha_1 \alpha_1^*}
{\big|[(k_1-k_2)^2+(k_1+k_2^*)^2]\alpha_1+(k_2-k_2^*-2k_1)(k_2+k_2^*)\beta_1\big|^2}}
\eea
and the expression for $T_2$ can be obtained by replacing $\alpha_1\leftrightarrow\beta_1$ and $\alpha_1^*\leftrightarrow\beta_1^*$ in the  expression for $T_1$.\\
\noindent\underline{Soliton $S_2$:}
\bea
q_1^{2+}\equiv q_2^{2+}=A^{2+} \mbox{sech}(\eta_{2R}^+) e^{i\eta_{2I}},
\eea
where $A^{2+}=\frac{(k_1-k_2)(k_1+k_2^*)}{(k_1^*-k_2^*)(k_1^*+k_2)}A^{2-}$ and $\eta_{2R}^+=\eta_{2R}
+\frac{\theta_{11}-\epsilon_{11}}{2}$. The real and imaginary parts of $\eta_{j}$-s appearing in the above expressions are already defined below equation (10).\\
\indent From the asymptotic analysis we observe that the amplitudes and hence the intensities before and after collision are not same for the non-degenerate soliton $S_1$, while it is so for $S_2$. It should be noticed that the arguments of the circular functions are also different before and after collision and also $L^{1-}\neq L^{1+}$. Additionally, there occurs a phase shift, $\Phi_1=\frac{\epsilon_{11}+R_2-\theta_{11}}{4k_{1R}}$, for soliton $S_1$. Thus, during its collision with soliton $S_2$, the soliton $S_1$ experiences an intensity switching among its two components resulting in a redistribution of the amplitude and phase. But the amplitude of the other soliton $S_2$ remains unaltered as $|A^{2+}|^2=|A^{2-}|^2$ and hence it undergoes the standard elastic collision only along with a phase shift $\Phi_2=\frac{\theta_{11}-\epsilon_{11}-R_2}{2k_{1R}}$. However, $S_2$ induces the collision with shape changes (intensity redistribution) in soliton $S_1$ during collision. 
\begin{figure}[h]
\centering
\includegraphics[width=0.6\linewidth]{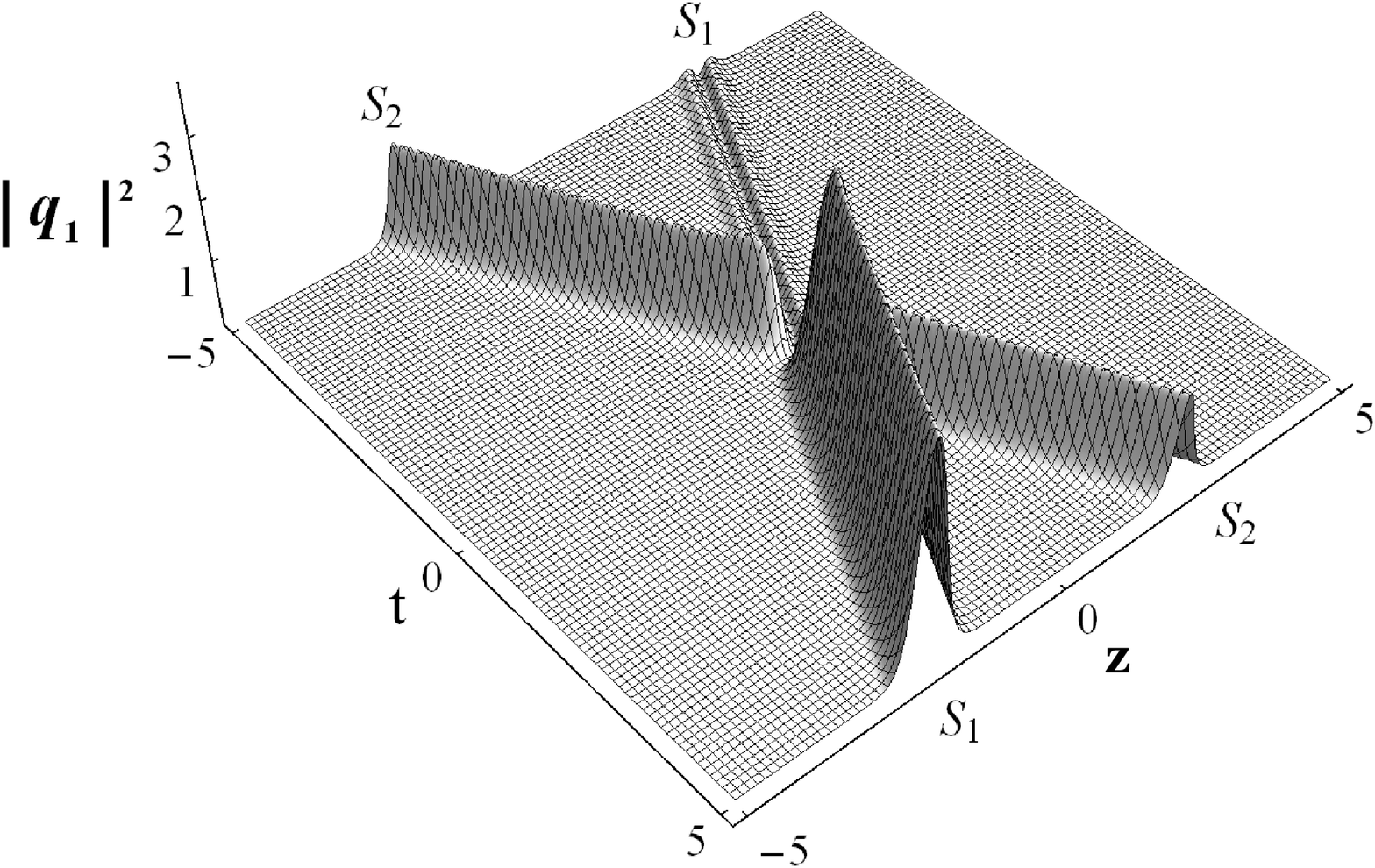}\\~~\includegraphics[width=0.6\linewidth]{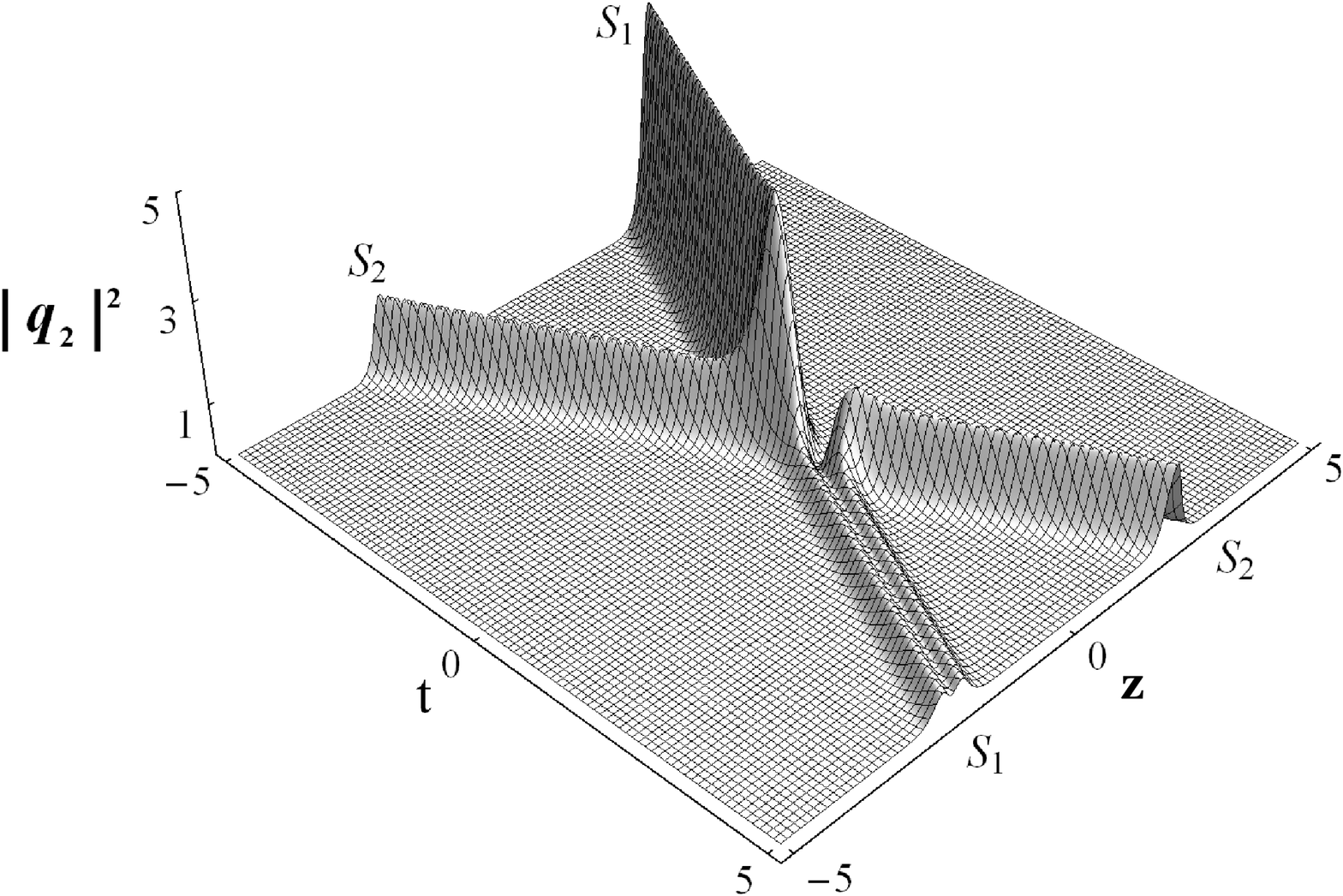}
\caption{Shape changing collision of a non-degenerate soliton $S_1$ with a degenerate soliton $S_2$ (parameters are as given in the text).}
\end{figure}
This collision scenario is quite different from the shape changing collision occurring in the Manakov system \cite{{r4},{r5},{r6}}, where there is an intensity redistribution among the solitons in both the components but in the present system it happens only among the two components of the non-degenerate soliton $S_1$. Note that though the total energy of both the solitons is conserved independently due to the conservation law,  $\int_{-\infty}^{ +\infty} (|q_1|^2+|q_2|^2)dt$=constant, the energy of the soliton in the individual modes  that is  $\int_{-\infty}^{ +\infty} |q_1|^2dt$ and $\int_{-\infty}^{ +\infty} |q_2|^2 dt$ are not conserved independently as the soliton $S_1$ only experiences intensity switching in both the components.  It could be an interesting future study to check whether $|T_j|$, $j=1,2$, can be unimodular, if so, for what choices of $\alpha$-s and $\beta$-s this will happen.\\
\indent For illustrative purpose, the above collision scenario is shown in figure 4 for the parameters $\gamma=2$, $k_1=2.3+i$, $k_2=2.5-i$, $\alpha_1=0.75$, $\beta_1=1.9$, and $\alpha_2=\beta_2=3+i$. The figure shows that in the $q_1$ component the single hump soliton $S_1$ changes its profile to a double hump soliton and also experiences significant suppression in its intensity whereas the soliton $S_2$ undergoes elastic collision. The reverse scenario takes place for the soliton $S_1$ in the $q_2$ component and here also the soliton $S_2$ remains unaltered after collision.\\
\indent In the collision of non-degenerate solitons alone the coherent coupling modifies uniformly both the solitons before and after collision, thereby resulting in an elastic collision. But in the present case the effect of coherent coupling is switched off in the degenerate soliton $S_2$ (since $\alpha_2^2=\beta_2^2$), however the coupling still persists in the non-degenerate soliton $S_1$. Hence, along with the XPM term the coherent coupling influences the non-degenerate soliton $S_1$ resulting in an intensity switching during collision. From a mathematical point of view, one finds that the asymptotically dominant terms of the non-degenerate soliton collision case become insignificant and the less dominant terms in that case become significant in the two-soliton solution expression corresponding to the collision of a degenerate soliton with a non-degenerate one. This yields different asymptotic expressions for these two collision processes as seen in the present subsection and in section 5.1, which ultimately makes their collision scenario completely different.

\section{Conclusion}
\indent Explicit forms of one- and two-soliton solutions of the coherently coupled NLS equations have been obtained using a non-standard type of Hirota's bilinearization method. Analysing the nature of the bright one-soliton solution we have reported degenerate solitons (solitons possessing same intensity in the $q_1$ and $q_2$ components) and non-degenerate solitons (solitons with different intensities in the $q_1$ and $q_2$ components). Particularly, for non-degenerate solitons the density profile can vary from single hump to double hump profile including flat-top solitons. Our analysis on the collision dynamics revealed the fact that separate collisions among degenerate solitons alone or among non-degenerate solitons alone are elastic. On the other hand, collision of a degenerate soliton with a non-degenerate soliton exhibits nontrivial behaviour resulting in an intensity switching of the non-degenerate soliton spread up in the two components leaving the other soliton unaltered. This property will have immediate applications in soliton collision based computing. Apart from the switching, we have also observed that this collision transforms the soliton profile from single hump to double hump including flat-top profile or vice versa. We expect that this property can find application in pulse shaping in the context of nonlinear optics.\\
\indent The above analysis can be extended to the study of three and higher order soliton solutions. The details of multi-soliton collisions and the multicomponent cases will be published separately.

\section*{\large{Acknowledgements}}
\indent TK acknowledges the support of the Department of Science and Technology, Government of India under the DST Fast Track Project for young scientists. TK also thanks the Principal and Management of Bishop Heber College, Tiruchirapalli, for constant support and encouragement. The works of MV and ML are supported by a DST-IRPHA project. ML is also supported by DST Ramanna Fellowship.

\section*{Appendix}
The various quantities occurring in equation (10) and in section 5 have the following forms:
\noindent\bea
\hspace{-2.5cm} e^{\delta _{ij}}=&&\frac{ \alpha _j^* (\alpha _i^2-\beta _i^2) \gamma }{2 (k_i+k_j^*)^2},\quad
e^{\delta _j}=\frac{ (\alpha _j^* (\alpha _1 \alpha _2-\beta _1 \beta _2) \gamma +(k_1-k_2) (\alpha _1 \kappa _{2j}-\alpha _2 \kappa_{1j}))}{(k_j+k_j^*) (k_{3-j}+k_j^*)},\nonumber\\
\hspace{-2.5cm} e^{\rho _{ij}}=&&-\frac{\beta_j^*}{\alpha_j^*}e^{\delta _{ij}}, \quad
e^{\rho _j}=\frac{ (\beta _j^* (-\alpha _1 \alpha _2+\beta _1 \beta _2) \gamma +(k_1-k_2) (\beta _1 \kappa _{2j}-\beta _2 \kappa_{1j}))}{(k_j+k_j^*) (k_{3-j}+k_j^*)},\nonumber\\
\hspace{-2.5cm} e^{\mu _{ij}}=&&\frac{ (k_1-k_2)^2 \alpha _{3-i}(\alpha _i^2-\beta _i^2) (\alpha _j^{*2}-\beta _j^{*2}) \gamma ^2}{4 (k_i+k_j^*)^4 (k_{3-i}^*+k_j)^2},\quad e^{\nu _{ij}}= \frac{\beta _{3-i}}{\alpha _{3-i}}e^{\mu _{ij}}\nonumber\\
\hspace{-2.5cm} e^{R_j}=&&\frac{ \kappa _{jj}}{(k_j+k_j^*)},~~ e^{\delta _0}=\frac{ \kappa _{12}}{(k_1+k_2^*)},~~~
 e^{\delta _0^*}=\frac{ \kappa _{21}}{(k_2+k_1^*)},~~\nonumber\\
\hspace{-2.5cm} e^{\phi _j}=&&\left(\frac{\gamma ^3 (k_1-k_2)^4 (k_1^*-k_2^*)^2 (\alpha _1^2-\beta _1^2)(\alpha _2^2-\beta_2^2)}
{8 (k_j+k_j^*)^4 (k_{3-j}+k_j^*)^4 (k_j+k_{3-j}^*)^2 (k_{3-j}+k_{3-j}^*)^2}\right)\alpha _{3-j}^*(\alpha _j^{*2}-\beta _j^{*2}),\nonumber\\
\hspace{-2.5cm} e^{\psi _j}=&&-\frac{\beta_{3-j}^*}{\alpha_{3-j}^*}e^{\phi _j},\quad 
e^{\epsilon _{ij}}=\frac{ \gamma ^2(\alpha _i^2-\beta _i^2) (\alpha _j^{*2}-\beta _j^{*2})}{4 (k_i+k_j^*)^4},\nonumber\\
\hspace{-2.5cm} e^{\tau _j}=&&\frac{\gamma ^2(\alpha _j^2-\beta _j^2) (\alpha _1^* \alpha _2^*-\beta _1^* \beta _2^*)}{2 (k_j+k_j^*)^2 (k_j+k_{3-j}^*)^2},\nonumber\quad
e^{\tau _j^*}=\frac{\gamma ^2(\alpha _j^{*2}-\beta _j^{*2}) (\alpha _1 \alpha _2-\beta _1 \beta _2) }{2 (k_j+k_j^*)^2 (k_j^*+k_{3-j})^2},\nonumber\\\
\hspace{-2.5cm} e^{\mu _1}=&&\frac{(k_1-k_2)^2 \gamma ^2(\alpha _1^2-\beta _1^2)}{\tilde{D}}
\left(\left[(k_2+k_1^*)^2+(k_2^*-k_1^*)(k_2^*+k_2)\right] \alpha _2 \alpha _1^* \alpha _2^*\right.\nonumber\\ 
\hspace{-2.5cm} &&\left.-(k_1^*-k_2^*) (k_2+k_2^*) \alpha _2^*\beta _2 \beta _1^* +
(k_2+k_1^*) (k_1^*-k_2^*) \alpha _1^*\beta _2 \beta _2^* -
(k_2+k_1^*) (k_2+k_2^*) \alpha _2 \beta _1^* \beta _2^* \right),\nonumber\\
\hspace{-2.5cm} e^{\mu _2}=&&\frac{(k_1-k_2)^2 \gamma ^2(\alpha _2^2-\beta _2^2)}{\tilde{D}}
\left([(k_1+k_1^*)^2+(k_2^*-k_1^*)(k_2^*+k_1)] \alpha _1 \alpha _1^* \alpha _2^*\right.\nonumber\\
\hspace{-2.5cm} &&\left.-(k_1^*-k_2^*) (k_1+k_2^*) \alpha _2^*\beta _1 \beta _1^* +
(k_1+k_1^*) (k_1^*-k_2^*) \alpha _1^*\beta _1 \beta _2^* -
(k_1+k_1^*) (k_1+k_2^*) \alpha _1 \beta _1^* \beta _2^* \right), \nonumber\\
\hspace{-2.5cm} e^{\nu _1}=&&\frac{-(k_1-k_2)^2 \gamma ^2(\alpha _1^2-\beta _1^2)}{\tilde{D}}
\left([(k_2+k_1^*)^2+(k_2^*-k_1^*)(k_2^*+k_2)] \beta _2 \beta _1^* \beta _2^*\right.\nonumber\\
\hspace{-2.5cm} && \left.-(k_1^*-k_2^*) (k_2+k_2^*) \beta _2^*\alpha _2 \alpha _1^* +
(k_2+k_1^*) (k_1^*-k_2^*) \beta _1^*\alpha _2 \alpha _2^* -
(k_2+k_1^*) (k_2+k_2^*) \beta _2 \alpha _1^* \alpha _2^* \right),\nonumber\\
\hspace{-2.5cm} e^{\nu _2}=&&-\frac{(k_1-k_2)^2 \gamma ^2(\alpha _2^2-\beta _2^2)}{\tilde{D}}
\left([(k_1+k_1^*)^2+(k_2^*-k_1^*)(k_2^*+k_1)] \beta _1 \beta _1^* \beta _2^*\right.\nonumber\\
\hspace{-2.5cm} &&\left.-(k_1^*-k_2^*) (k_1+k_2^*) \beta _2^*\alpha _1 \alpha _1^* +
(k_1+k_1^*) (k_1^*-k_2^*) \beta _1^*\alpha _1 \alpha _2^* -
(k_1+k_1^*) (k_1+k_2^*) \beta _1 \alpha _1^* \alpha _2^*\right), \nonumber\\
\hspace{-2.5cm} e^{R_3}=&&\gamma ^2\Big[\left(\frac{(k_1^*+k_2)^2 (k_1+k_2^*)^2-(k_1+k_1^*) (k_1^*+k_2) (k_1+k_2^*) (k_2+k_2^*)+(k_1+k_1^*)^2 (k_2+k_2^*)^2}{(k_1+k_1^*)^2
(k_1^*+k_2)^2 (k_1+k_2^*)^2 (k_2+k_2^*)^2}\right)\nonumber\\
\hspace{-2.5cm} && (\alpha _1 \alpha _1^* \alpha _2 \alpha _2^* +\beta _1 \beta _1^* \beta _2 \beta _2^* )+
\left(\frac{(k_1-k_2) (k_1^*-k_2^*) (\alpha _2 \alpha _2^* \beta _1 \beta _1^*+\alpha _1 \alpha _1^* \beta _2 \beta _2^* )}{(k_1+k_1^*)^2
(k_1^*+k_2) (k_1+k_2^*) (k_2+k_2^*)^2}\right)\nonumber\\
\hspace{-2.5cm} &&  -\left(\frac{(\alpha _1^* \alpha _2^*\beta _1 \beta _2 +\alpha _1 \alpha _2\beta _1^* \beta _2^* )}{(k_1+k_1^*)
(k_1^*+k_2) (k_1+k_2^*) (k_2+k_2^*)}\right) -
\left(\frac{(k_1-k_2) (k_1^*-k_2^*)(\alpha _1 \alpha _2^* \beta _2 \beta _1^*+\alpha _2 \alpha _1^* \beta _1 \beta _2^* )}{(k_1+k_1^*) (k_1^*+k_2)^2
(k_1+k_2^*)^2 (k_2+k_2^*)}\right)\Big],\nonumber
\eea
\bea
\hspace{-2.5cm} &&e^{\theta _{ij}}=\frac{(k_1-k_2)^2 (k_1^*-k_2^*)^2 (\alpha _i^2-\beta _i^2)(\alpha _j^{*2}-\beta _j^{*2}) (\alpha _{3-i} \alpha _{3-j}^*+\beta_{3-i} \beta _{3-j}^*) \gamma ^3 }{{2\tilde{D}}(k_i+k_j^*)^2},\nonumber\\
\hspace{-2.5cm} &&e^{R_4}=\frac{1}{4\tilde{D}^2}(k_1-k_2)^4 (k_1^*-k_2^*)^4 (\alpha _1^2-\beta _1^2) (\alpha _1^{*2}-\beta _1^{*2}) (\alpha _2^2-\beta _2^2) (\alpha _2^{*2}-\beta_2^{*2}) \gamma ^4,\nonumber\\
\hspace{-2.5cm} &&e^{\lambda _{ij}}=\frac{ (k_1-k_2)^2 (\alpha _{3-i}^2-\beta _{3-i}^2)\kappa_{ij}\gamma }{(k_{3-i}+k_j^*)^2(k_i+k_j^*)},\quad
e^{\lambda _j}=\frac{\gamma ^2(k_1-k_2)^4 (\alpha _1^2-\beta _1^2) (\alpha _2^2-\beta _2^2)(\alpha _j^{*2}-\beta _j^{*2})\gamma ^2}{4 (k_j+k_j^*)^4 (k_j^*+k_{3-j})^4},\nonumber\\
\hspace{-2.5cm} &&e^{\lambda _3}=\frac{1}{\tilde{D}}(k_1-k_2)^4 (\alpha _1^2-\beta _1^2) (\alpha _2^2-\beta _2^2) (\alpha _1^* \alpha _2^*-\beta _1^* \beta _2^*)\gamma^2, \nonumber
\eea
where
\bea
\tilde{D}=2 (k_1+k_1^*)^2(k_1^*+k_2)^2 (k_1+k_2^*)^2 (k_2+k_2^*)^2 \nonumber
\eea
and
\bea
\kappa_{ij}=\frac{\gamma (\alpha_i \alpha_j^*+\beta_i \beta_j^*)}{(k_i+k_j^*)},\quad i, j=1,2. \nonumber
\eea\endnumparts

\end{document}